\documentclass[structabstract,longauth]{aa}  
\usepackage{graphicx}
\usepackage{txfonts}
\usepackage{natbib}
\bibpunct{(}{)}{;}{a}{}{,} 
\usepackage{color}
\newcommand{\cmtwo}{cm$^{-2}$}  
\newcommand{\cmthree}{cm$^{-3}$}

\newcommand{\kms}{km\,s$^{-1}$}       
\newcommand{\ms}{m\,s$^{-1}$} 
\newcommand{\cms}{cm\,s$^{-1}$} 
\newcommand{\vlsr}{$\upsilon_{\rm LSR}$}        

\newcommand{\ta}{$T_{\rm A}$}        

\newcommand{\tadv}{$\int \!\! T_{\rm A} {\rm d}\upsilon$}

\newcommand{\tmbdv}{$\int \!\! T_{\rm mb}\,{\rm d}\upsilon$}
\newcommand{\tmb}{$T_{\rm mb}$}

\newcommand{\um}{$\mu$m}                                 
\newcommand{\molh}{H$_{2}$}                              

\newcommand{\water}{H$_{2}$O}

\newcommand{\molo}{O$_{2}$}			
\newcommand{\moloiso}{${\rm O}^{18}{\rm O}$}

\newcommand{\gapprox}{$\stackrel {>}{_{\sim}}$}   
\newcommand{\lapprox}{$\stackrel {<}{_{\sim}}$}
\newcommand{\about}{$\sim$}                       
\newcommand{\powten}[1]{10$^{#1}$}
\newcommand{\av}{$A_{\rm V}$}                     

\newcommand{\ro}{$\rho \, {\rm Oph}$}
\newcommand{\roc}{$\rho \, {\rm Oph \,\, cloud}$}
\newcommand{\roa}{$\rho \, {\rm Oph \, A}$}

\newcommand{\roac}{$\rho \, {\rm Oph \, A \, core}$}

\newcommand{\amin}{$^{\prime}$}                   
\newcommand{\asec}{$^{\prime \prime}$}
\newcommand{\adeg}{$^{\circ}$}

\newcommand{\radot}[4]{\mbox{#1$^{\rm h}$#2$^{\rm m}$#3$\stackrel {\rm s}{_{\bf\cdot}}$#4}}  
\newcommand{\decdms}[3]{\mbox{#1$^{\circ}$#2$^{\prime}$#3$^{\prime \prime}$}}

\newcommand{\asecdot}[2]{\mbox{#1$\stackrel {\prime \prime}{_{\bf \cdot}}$#2}}

\begin{document}
   \title{Multi-line detection of \molo\  toward \roa\thanks{Based on observations with {\it Herschel}-HIFI. {\it Herschel} is an ESA space observatory with science instruments provided by European-led Principal Investigator consortia and with important participation from NASA. } }


   \author{R. Liseau
          \inst{1}
          \and
          P.F. Goldsmith\inst{2}
          \and
          B. Larsson\inst{3} 
          \and
          L. Pagani\inst{4}
          \and
          P. Bergman\inst{5}
          \and
          J. Le Bourlot\inst{6}
          \and
          T.A. Bell\inst{7}
          \and
          A.O. Benz\inst{8}
          \and
          E.A. Bergin\inst{9}
          \and
          P. Bjerkeli\inst{1}
          \and
          J.H. Black\inst{1}
          \and
          S. Bruderer\inst{8,\,21}
          \and
          P. Caselli\inst{10}
          \and
          E. Caux\inst{11}
          \and
          J.-H. Chen\inst{2}
          \and
          M. de Luca\inst{6}
          \and
          P. Encrenaz\inst{4}
          \and
          E. Falgarone\inst{12}
          \and
          M. Gerin\inst{12}
          \and
          J.R. Goicoechea\inst{7}
          \and
          \AA. Hjalmarson\inst{1}
          \and
          D.J. Hollenbach\inst{13}        
          \and
          K. Justtanont\inst{1}
          \and
          M.J. Kaufman\inst{14}
          \and
          F. Le Petit\inst{6}
          \and
          D. Li\inst{15,\,16}
          \and
          D.C. Lis\inst{16}
          \and
          G.J. Melnick\inst{17}
          \and
          Z. Nagy\inst{18}
          \and
	A.O.H. Olofsson\inst{5}
          \and
          G. Olofsson\inst{3}
          \and
          E. Roueff\inst{6}
          \and
          Aa. Sandqvist\inst{3}
          \and
          R.L. Snell\inst{19}
          \and
          F.F.S. van der Tak\inst{18}
          \and
          E.F. van Dishoeck\inst{20,\,21}
          \and
          C. Vastel\inst{11}
          \and
          S. Viti\inst{22}
          \and
          U.A. Y{\i}ld{\i}z\inst{20}          
          }

   \institute{Department of Earth and Space Sciences, Chalmers University of Technology, Onsala Space Observatory, SE-439 92 Onsala, Sweden,
              \email{rene.liseau@chalmers.se}
         \and
             	Jet Propulsion Laboratory, California Institute of Technology, 4800 Oak Grove Drive, Pasadena CA 91109, USA 
	\and
		Department of Astronomy, Stockholm University, SE-106 91 Stockholm, Sweden
	\and 
		LERMA \& UMR8112 du CNRS, Observatoire de Paris, 61 Av. de l'Observatoire, 75014, Paris, France
	\and	
		Onsala Space Observatory, Chalmers University of Technology, SE-439 92 Onsala, Sweden
	\and	
		Observatoire de Paris, LUTH, Paris, France
	\and
		Centro de Astrobiolog{\'i}a, CSICÐINTA, 28850, Madrid, Spain
	\and 
		Institute of Astronomy, ETH-Zurich, Zurich, Switzerland
	\and
		Department of Astronomy, University of Michigan, 500 Church Street, Ann Arbor MI 48109, USA
	\and
		School of Physics and Astronomy, University of Leeds, Leeds, UK
	\and
		Universit\'e de Toulouse; UPS-OMP; IRAP; Toulouse, France \& CNRS; IRAP; 9 Av. colonel Roche, BP 44346, F-31028 Toulouse Cedex 4, France		
	\and
		LRA/LERMA, CNRS, UMR8112, Observatoire de Paris \& \'Ecole Normale Sup\'erieure, 24 rue Lhomond, 75231 Paris Cedex 05, France
	\and
		SETI Institute, Mountain View CA 94043, USA
	\and
		Department of Physics and Astronomy, San Jos\'e State University, San Jose CA 95192, USA
	\and
		National Astronomical Observatories, Chinese Academy of Sciences, A20 Datun Road, Chaoyang District, Beijing 100012, China
	\and
		California Institute of Technology, Cahill Center for Astronomy and Astrophysics 301-17, 1200 E. California Boulevard, Pasadena CA 91125, USA
	\and
		Harvard-Smithsonian Center for Astrophysics, 60 Garden Street, MS 66, Cambridge MA 02138, USA
	\and
		SRON Netherlands Institute for Space Research, PO Box 800, 9700 AV, and Kapteyn Astronomical Institute, University of Groningen, Groningen, The Netherlands
	\and
		Department of Astronomy, University of Massachusetts, Amherst MA 01003, USA
	\and
		Leiden Observatory, Leiden University, PO Box 9513, 2300 RA Leiden, The Netherlands
	\and
		Max-Planck-Institut f\"ur Extraterrestrische Physik, Gie\ss enbachstra\ss e 1, 85748 Garching, Germany		
	\and
		Department of Physics and Astronomy, University College London, London, UK
             }

   \date{Received ; accepted }

 
  \abstract
   {Models of pure gas-phase chemistry in well-shielded regions of molecular clouds predict relatively high levels of molecular oxygen, \molo, and water, \water. These high abundances would imply large cooling rates, leading to relatively short time scales for the evolution of gravitationally unstable dense cores, forming stars and planets. Contrary to expectation, the dedicated space missions SWAS and Odin found typically only very small amounts of water vapour and essentially no \molo\ in the dense star-forming interstellar medium. }
   {Only toward \roa\ did Odin detect a very weak line of \molo\ at 119\,GHz in a beam size of 10 arcmin. Line emission of related molecules changes on angular scales of the order of some tens of arcseconds, requiring a larger telescope aperture such as that of  the {\emph {Herschel}} Space Observatory to resolve the \molo\ emission and to pinpoint its origin. }
   {We use the Heterodyne Instrument for the Far Infrared (HIFI) aboard {\emph {Herschel}}  to obtain high resolution \molo\ spectra toward selected positions in the \roac. These data are analysed using standard techniques for \molo\ excitation and compared to recent PDR-like chemical cloud models.}
   {The $N_J =  3_3 - 1_2$ line at  487.2\,GHz was clearly detected toward all three observed positions in the  \roac. In addition, an oversampled map of the $5_4-3_4$ transition at 773.8\,GHz revealed the detection of the line in only half of the observed area. Based on their ratios, the temperature of the \molo\ emitting gas appears to vary quite substantially, with warm gas (\gapprox\,$50$\,K) adjacent to a much colder region, where temperatures are below 30\,K.}
   { The exploited models predict \molo\ column densities to be sensitive to the prevailing dust temperatures, but rather insensitive to the temperatures of the gas. In agreement with these model, the observationally determined \molo\ column densities seem not to depend strongly on the derived gas temperatures, but fall into the range $N$(\molo) = 3 to \gapprox\,$6 \times 10^{15}$\,\cmtwo. Beam averaged \molo\ abundances are about $5 \times 10^{-8}$ relative to \molh. Combining the HIFI data with earlier Odin observations yields a source size at 119\,GHz of about 4 to 5\,arcmin, encompassing the entire \roac. We speculate that the general very low detection rate of \molo\ sources might be explained by the transient nature of interstellar \molo\ molecules.}
   
               
   \maketitle
%

\section{Introduction}

Despite the universal importance of oxygen, its molecular form, \molo, is an elusive species of the interstellar medium (ISM). This molecule was thought to be one of the main regulators of the energy balance in the ISM, as summarised by \citet{goldsmithlanger1978}. Consequently, large efforts, both from ground (\moloiso) and space, had been devoted to obtain quantitative estimates of its abundance. The historical account of this essentially fruitless ``\molo-struggle'' was recently reviewed by \citet{goldsmith2011}. Prior to {\it Herschel}, both SWAS \citep{melnick2000,goldsmith2000} and Odin \citep{nordh2003,pagani2003} had already shown that 
abundances for  both \molo\ and \water\ assumed by  \citet{goldsmithlanger1978} were much larger than actual values in the ISM and, with the exception of the \roc, none of the \molo\ lines were detected anywhere in the ISM. \citet{goldsmith2002} announced the tentative detection by SWAS of the \molo\,487\,GHz line in \roa. The claimed signal appeared at an unusual velocity and was atypically broad. \citet{pagani2003} showed that, on the basis of more sensitive \molo\ 119\,GHz observations with Odin, this was an erroneous result. This line was finally detected by Odin, at the correct LSR-velocity and with a plausible, narrow line width \citep{larsson2007}. Here we report the detection of two more \molo\ transitions in \roa, viz. at 487\,GHz and at 774\,GHz, respectively.

Besides being relatively nearby \citep[120--130\,pc,][]{lombardi2008,snow2008,mamajek2008,loinard2008} the \roc\ distinguishes itself from other low-mass star forming regions in that it exhibits evidence (e.g., in C$^{18}$O line emission) of gas at relatively high temperatures ($T \ge 20$\,K) over extended regions with high column densities \citep[$N({\rm H}_2) \gg 10^{22}$\,\cmtwo,][]{liseau2010}. In addition, \roa\ displays an interesting chemistry: For example, doubly deuterated formaldehyde (D$_2$CO) and  hydrogen peroxide (H$_2$O$_2$) molecules, rarely seen elsewhere in the ISM have been found here (Bergman et al. 2011a, b). The overall impression is that the observable abundance of many species is the result of surface reactions on dust grains, a process which may also pertain to the production of oxygen molecules. 

\begin{figure*}[ht]
  \resizebox{\hsize}{!}{
  \rotatebox{90}{\includegraphics{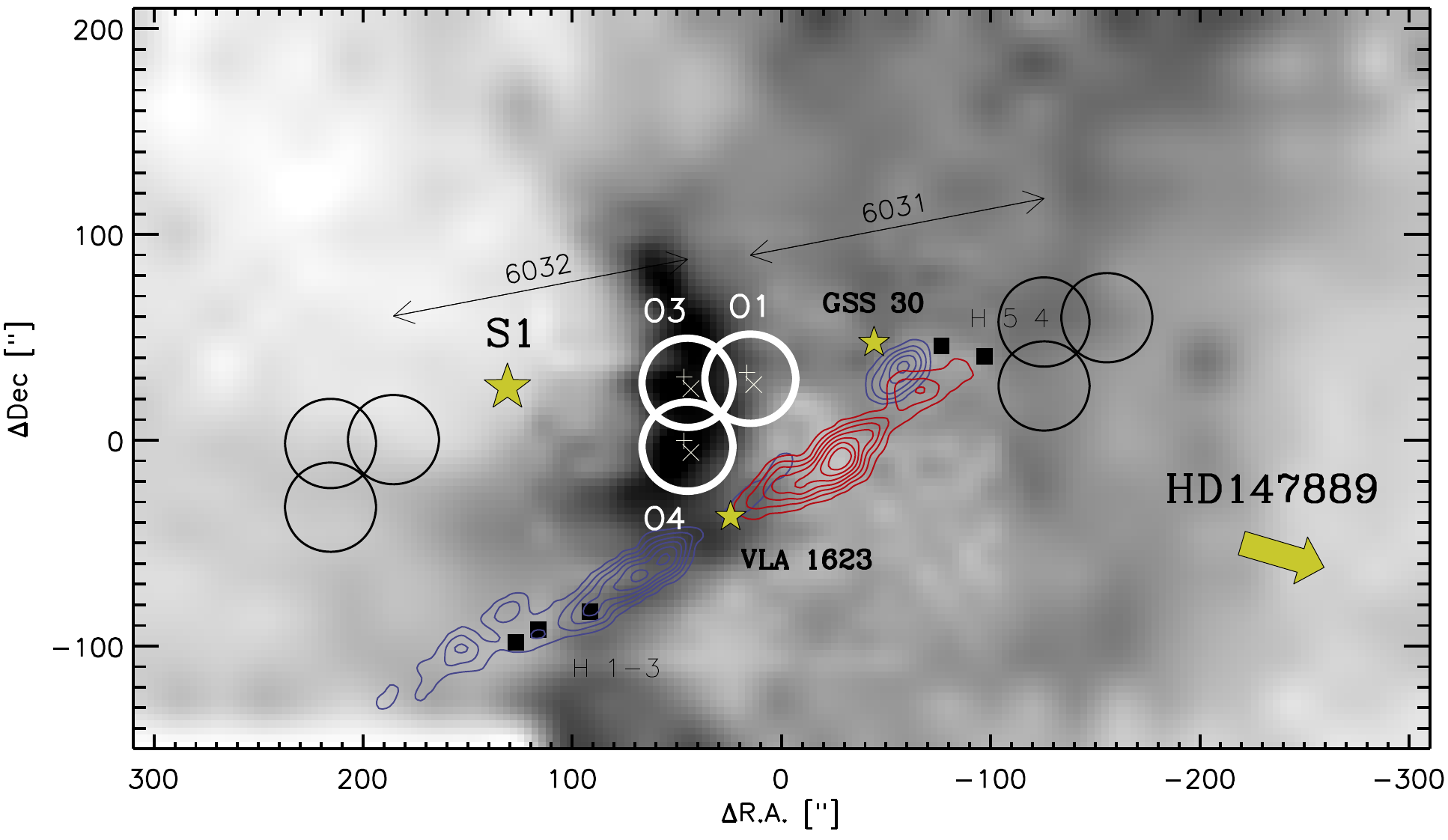}}
                        }
  \caption{Herschel beams of \molo\,487\,GHz (44\asec) superposed onto a grey scale C$^{18}$O\,(3-2) map of \roa\ \citep{liseau2010}. The $+$ and $\times$ symbols designate the positions of the H- and V-polarization beam centers of HIFI, respectively, with a separation of \asecdot{6}{6}. The heavy circles show the positions of the sources O1, O3 and O4 (see Table\,\ref{sources}), with the corresponding off-source reference positions  shown by the displaced  symbols to the east and west, respectively, and identified by their observation numbers.  The origin is at \radot{16}{26}{24}{6}, \decdms{$-24$}{23}{54} J(2000.0) and offsets in R.A. and Dec. are in arcsec.  At the distance of 120\,pc, 40\asec\ corresponds to about 0.02\,pc. The star symbols identify the positions of the star S\,1, the young stellar object GSS\,30 and the Class\,0 source VLA\,1623, where the contours depict red- and blueshifted CO\,(3--2) emission of its bipolar outflow. The filled squares identify the known Herbig-Haro objects H1 through 5 and the direction to the dominating B-star HD\,147889, about 15\amin\ away, is indicated by the heavy arrow. }
  \label{dbs}
\end{figure*}

\begin{table*}
\caption{\molo\ transitions and instrumental characteristics.}             
\label{lines}      
\begin{tabular}{cccccccl}      
\hline\hline    
\noalign{\smallskip}             
Transition$^a$ 	& Frequency$^b$	& Wavelength	 		& Energy 			& Observatory/		& Beam width  		& Efficiency		&Reference \\
$N_J^{\rm up}-N_J^{\rm low}$		&  $\nu_0$ (MHz)	& $\lambda_0$ (\um)	& $E_{\rm up}/k$ (K)	& Instrument		& HPBW (\amin)	& $\eta_{\rm mb}$	&	\\
\noalign{\smallskip}	
\hline                        
\noalign{\smallskip}	
$1_1-1_0$	 & 118750.341		& 2526				&  \phantom{1}5.7	&	Odin			& 9.96			&0.91			 & \citet{frisk2003}	\\
$3_3-1_2$	 & 487249.264		& \phantom{1}616		& 26.4			&	SWAS		& $3.5\times 5.0$	&0.90			 & \citet{melnick2000}	\\
			 & 				& 					& 				&	HIFI			& 0.73			& \phantom{1}0.757	&  \citet{roelfsema2012}	\\
$5_4-3_4$	 & 773839.512		&  \phantom{1}388		& 60.7			&	HIFI			& 0.47			& \phantom{1}0.753	 &  \citet{roelfsema2012} \\
\hline                                   
\end{tabular}
\begin{list}{}{}
\item[$^{a}$] An energy level diagram is provided by, e.g., \citet{goldsmith2011}.
\item[$^{b}$] Frequencies are from the JPL-catalogue \texttt{http://spec.jpl.nasa.gov/cgi-bin/catform}.
\end{list}
\end{table*}

The motivation for the {\it Herschel} observations was, firstly, to pin down the precise location of the \molo\ source inside the large, ten-arcminute beam of Odin. Secondly,  to add observations of the \molo\ 487.25\,GHz ($3_3$-$1_2$) and 773.84 GHz ($5_4$-$3_4$) lines, which have upper level energies $E_{\rm up}/k=26$\,K and 61\,K, respectively, to the Odin data of the 118.75 GHz ($1_1$-$1_0$) transition ($E_{\rm up}/k=6$\,K).
These observations should enable us to learn about the nature of the \molo\ source in the \roc. This would then be compared to the physical and chemical conditions of other locations in the general ISM and potentially identify the characteristics and time scales of regions containing \molo\ molecules. 

The paper is organised as follows: in Sect.\,2, our Herschel-HIFI observations and their reduction are discussed in considerable detail and our results are presented in Sect.\,3. These results are discussed in Sect.\,4, where the spectral line formation is analysed under different assumptions. Finally, in Sect.\,5, our main conclusions are briefly summarised. 

\begin{figure*}[ht]
  \resizebox{\hsize}{!}{
  \rotatebox{00}{\includegraphics{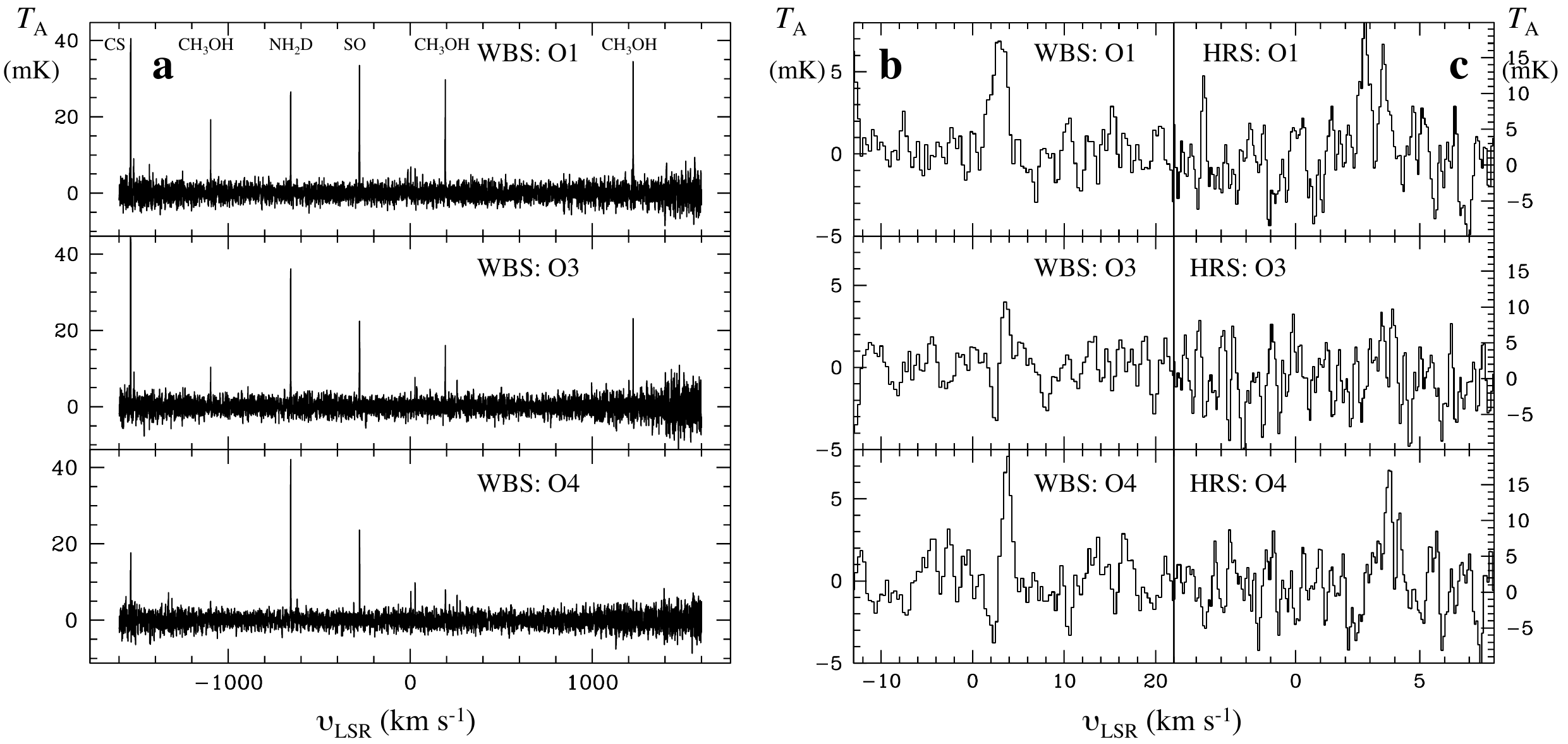}}
                        }
  \caption{Averaged Double Side Band (DSB) spectra of H- and V-polarizations for the three positions O1, O3 and O4 in the \roa\ cloud core, shown for the range 484.65\,GHz to 489.85\,GHz. The spectral lines are found exclusively in the Lower Side Band (LSB) and the LSR velocity, \vlsr, along the abcissa is in \kms\ and the antenna temperature \ta\ along the ordinate in mK. {\bf a:} 5\,GHz wide band of the WBS around the 487\,\molo\ line ($-1600,\,+1600$\,\kms). {\bf b:}  The WBS spectrum at expanded scale, showing the \molo\ line at 487\,GHz, which is detected in all three positions. Inspection of the individual on-off pairs suggests that the apparent absorption features next to the lines are not real but simply due to the noise. {\bf c:} Similar to {\bf b} but for the corresponding HRS data.}
  \label{obs}
\end{figure*}

\begin{table*}
\caption{Source designations and coordinates for \roa, with positions observed in \molo\ in bold face.}             
\label{sources}      
\centering                          
\begin{tabular}{l l l l}      
\hline\hline    
\smallskip             
RA (h m s) J2000.0	 &	Dec (\adeg\,\amin\,\asec) J2000.0	& Source Designation &	Reference \\
\hline                        
16 25 24.32	 & $-24$ 27 56.57	& HD\,147889			& SIMBAD: http://simbad.u-strasbg.fr/simbad/\\
16 26 17.5	 & $-24$ 23 13		& H 4, HH 313B 		& \citet{dent1995}, \citet{caratti2006}	\\
16 26 19.0	 & $-24$ 23 08		& H 5, HH 313A		& \citet{dent1995}, \citet{caratti2006}	\\
16 26 21.36	 & $-24$ 23 06.4	& GSS 30, El 21		& \citet{grasdalen1973}, \citet{elias1978}\\
{\bf 16 26 25.7}	 & {\bf --24 23 24}	& {\bf O1}				& this paper \\
{\bf 16 26 25.7}	 & {\bf --24 23 57}	& {\bf O2}				& this paper \\
16 26 26.1	 & $-24$ 23 14 		& N$_2$H$^+$ N1\,b (FWHM=0.29\,\kms) & \citet{diFrancesco2004} \\
16 26 26.38 	 & $-24$ 24 31.0	& VLA\,1623	 		& \citet{andre1990}\\
16 26 27.1	 & $-24$ 23 30		& N5, 850\,\um 			& \citet{diFrancesco2004}, \citet{johnstone2000} \\
16 26 27.2  	 & $-24$ 24 04	 	& Deuterium peak		& \citet{bergman2011a} \\
16 26 27.3	 & $-24$ 23 28		& SM\,1N, 1.3\,mm  		& \citet{motte1998} \\
{\bf 16 26 27.9}	 & {\bf --24 23 26}	& {\bf O3}, P2,  C$^{18}$O (3 - 2) 	& this paper, \citet{liseau2010} \\
{\bf 16 26 27.9}	 & {\bf --24 23 57}	& {\bf O4}, P3,  C$^{18}$O (3 - 2); SM\,1, 1.3\,mm &  this paper, \citet{liseau2010}, \citet{motte1998} \\
16 26 34.19	 & $-24$ 23 28.2	& S 1, GSS 35, El 25 	& \citet{grasdalen1973}, \citet{elias1978}\\
\hline                                   
\end{tabular}
\end{table*}

\begin{figure}[ht]
  \resizebox{\hsize}{!}{
  \rotatebox{0}{\includegraphics{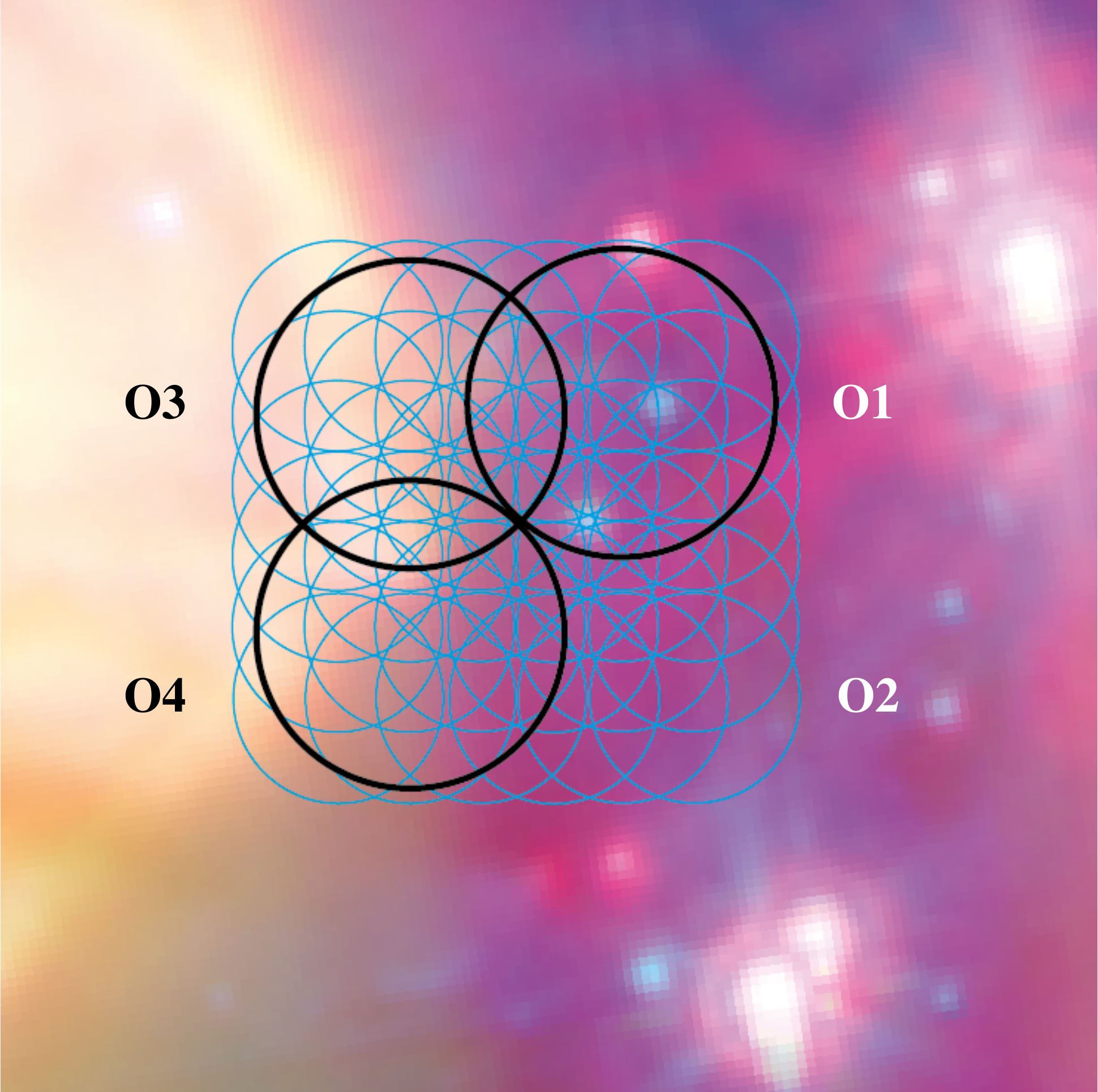}}
                        }
  \caption{Outline of the oversampled $6 \times 6$ regular map with 10\asec\ spacings of the \molo\,774\,GHz observations. At this frequency, the beam of the 3.5\,m Herschel telescope has a size of 28\asec\ (small blue circles). The beam centers for the H- and V-polarisations are separated by \asecdot{4}{5}. Larger (black) circles correspond to the 487\,GHz beam positions (44\asec; cf. Fig\,\ref{dbs}) and as before, North is up and East is to the left. Background image is courtesy of {\it Spitzer}-IRAC-MIPS (blue: 3.6\,\um, green: 8\,\um, red: 24\,\um).}
  \label{774_obs}
\end{figure}

\section{Observations and data reduction}

{\it Herschel} is a space platform for far infrared and sub-millimetre observations \citep{pilbratt2010}. It is orbiting the Sun about 1.5 million kilometres beyond the Earth (at L\,2) and its 3.5\,m primary dish is radiatively cooled to its operational equilibrium temperature of about 85\,K. The scientific instruments are placed in a liquid helium filled cryostat, limiting its cold lifetime to roughly 3.5 years. One of the three onboard instruments is the Heterodyne Instrument for the Far Infrared \citep[HIFI, ][]{degraauw2010} with continuous frequency coverage from 480 to 1250\,GHz ($\lambda \lambda\,624 - 234$\,\um) in 5 bands. In addition, the frequency range 1410 to 1910\,GHz ($\lambda \lambda\,213 - 157$\,\um) is covered by bands 6 and 7. The spectral resolving capability is the highest available on {\it Herschel}, i.e. up to \powten{7} corresponding to 30\,\ms, required to resolve the profiles of very narrow lines of cold sources, having temperatures of the order of 10\,K or less. 

Spectral line data and relevant characteristics of the instruments discussed in this paper have been compiled in Table\,\ref{lines}.

\subsection{487\,GHz observations}

The 487\,GHz observations with HIFI were performed on operating day OD\,673 (2011 March 18). The three positions O1, O3 and O4 at the centre of the \roac\ (Table\,\ref{sources}) were observed in double beam-switch mode (DBS), with beam throws of 3\arcmin\ in roughly the  west and east directions (with a position angle of 101\adeg\ E of N) and identified in Fig.\,\ref{dbs} by the numbers 6031 and 6032, respectively. For each position, the relative observing time spent on the east- and westward  on-off pairs was 7.3\,hr, and the total programme execution time was 21.9\,hr. The DBS mode generally produces the highest quality data with HIFI and since the \molo\ signal is known to be weak, this was the observing mode chosen. Switching by only three arcminutes inside a molecular cloud is generally considered not a sensible option, as this potentially results in cancellation of the source signal. However, a relatively large velocity gradient (\gapprox\,0.3\,\kms/arcmin) and/or two distinctly different velocity systems ($\Delta \upsilon \sim 1$\,\kms) are known to exist in \roa\ \citep[e.g.,][]{bergman2011a}. In addition, the \molo\ line is known to be narrow (${\rm FWHM} \sim 1$\,\kms), which taken together largely diminish the risk of signal cancellation. In fact, splitting the entire data set into two and reducing these two halves with only one off-spectrum used resulted in essentially the same spectral data. These off-beams were pointing at very different regions in \roa\ (Fig.\,\ref{dbs}), but no traceable off-beam contamination was introduced by the DBS observations.

\begin{table}
\caption{\molo\ line observations of \roa.}             
\label{results}      
\centering                          
\begin{tabular}{lcccc}      
\hline\hline    
\noalign{\smallskip}             
Position/		 	&	$T_{\rm mb,\,0}^{a}$ &	 \vlsr$^{a}$		&	FWHM$^{a}$	&	$10^3$\tmbdv	 \\
Average			&	(mK)				&	(\kms)			&	(\kms)		& 	(K km s$^{-1}$)	 \\	
\noalign{\smallskip}     
\hline	                       
\noalign{\smallskip}   
				&	{\bf 487\,GHz}		&$N_J=3_{3}$-$1_{2}$	&				&				\\       
\noalign{\smallskip} 
\hline                                 
\noalign{\smallskip}
O1 HRS			&	 $ 14.4 \pm 1.8$	&	$3.05 \pm 0.10$	& $1.63 \pm 0.23$	& $19.8 \pm 2.4$	\\
\hspace*{0.5cm}O1$\_1^{b}$ &$ 24.2 \pm 3.3$	&	$2.79 \pm 0.05$	& $0.46 \pm 0.08$	& $15.5 \pm 1.7$	\\
\hspace*{0.5cm}O1$\_2$ &$ 19.6 \pm 3.5$	&	$3.58 \pm 0.04$	& $0.41 \pm 0.09$	&\phantom{1}$8.7\pm 1.7$\\
O3 HRS			&\phantom{1}$7.6\pm 2.2$&	$3.59 \pm 0.12$	& $0.80 \pm 0.27$	&\phantom{1}$5.7 \pm 1.3$\\	
O4 HRS			& $ 11.5\pm 2.5$		&	$3.75 \pm 0.07$	& $0.84 \pm 0.16$	& $13.0 \pm 1.6$     \\
\noalign{\smallskip}
O1 WBS			&\phantom{1}$9.4 \pm 2.0$&	$2.89 \pm 0.20$	& $1.91 \pm 0.46$	& $18.4\pm 1.6$	\\
\hspace*{0.5cm}O1$\_1$ &\phantom{1}$ 9.6 \pm 1.4$&	$2.70$ fix		& $0.90$ fix		&\phantom{1}$9.2 \pm 1.4$\\
\hspace*{0.5cm}O1$\_2$ &$\phantom{1}6.9 \pm 1.4$&	$3.60$ fix		& $0.90$ fix		&\phantom{1}$6.6 \pm 1.4$\\
O3 WBS			&\phantom{1}$5.7 \pm 3.4$&	$3.66 \pm 0.26$	& $0.87 \pm 0.60$	&\phantom{1}$5.3 \pm 1.5$\\
O4 WBS			&	$10.3 \pm 2.6$		&	$3.78 \pm 0.12$	& $0.99 \pm 0.28$	& $10.3 \pm 1.5$ \\
\noalign{\smallskip}
$<$O3, O4$>$WBS 	&\phantom{1}$7.8 \pm 3.0$&	$3.69 \pm 0.20$	& $0.90 \pm 0.47$	&\phantom{1}$7.8 \pm 1.5$	\\	
\hline  
\noalign{\smallskip} 
				&	\bf {774\,GHz}		&$N_J=5_{4}$-$3_{4}$	&				&				\\   
\noalign{\smallskip}          
\hline                                 
\noalign{\smallskip}
$<$O1, O2$>$HRS 	&	$22.3 \pm 9.9$		&	$3.59 \pm 0.03$	& $0.57 \pm 0.10$	& $13.2 \pm 4.0$	\\
O1 WBS			&	$17.9 \pm 4.2$		&	$3.66 \pm 0.07$	& $0.61 \pm 0.16$	& $11.4 \pm 2.9$	\\
O2 WBS			&       $14.5 \pm 3.8$		&	$3.59 \pm0.09$	& $0.72 \pm 0.22$	& $10.9 \pm 2.8$	\\
$<$O1, O2$>$WBS 	&	$15.9 \pm 2.8$		&	$3.62 \pm 0.06$	& $0.67 \pm 0.17$	& $11.2 \pm 2.0$	\\
\noalign{\smallskip}
O3 WBS			&	$\dots$			&	$\dots$			& $\dots$			&\phantom{1}$3.6 \pm 2.9$	\\
O4 WBS			&	$\dots$			&	$\dots$			& $\dots$			& $-1.6 \pm 3.0$	\\
$<$O3, O4$>$WBS 	&	$\dots$			&	$\dots$			& $\dots$			&\phantom{1}$0.8 \pm 2.0$	\\	
\hline
\noalign{\smallskip} 
				&	\bf {119\,GHz}		&$N_J=1_1$-$1_0$	&				&				\\   
\noalign{\smallskip}          
\hline                                 
\noalign{\smallskip}
Odin$^{c}$		&	$19.3 \pm 0.6$		&	$3.5 \pm 0.5$		& $1.5 \pm 0.5$	& $26.7 \pm 4.4$	\\
\hline                                 
\end{tabular}
\begin{list}{}{}
\item[$^{a}$] Gaussian fits: peak intensity, central velocity and full width half maximum of line.
\item[$^{b}$] O1$\_1$ and O1$\_2$ refer to two velocity components (see Sect.\,3.1).
\item[$^{c}$] 10\amin\ beam centered on \roa\ \citep{larsson2007}.
\end{list}
\end{table}

The receivers are double side band and to counteract confusion in frequency space, we had selected for each observation 8 different LO-tunings inside HIFI-band\,1a, spaced by 170\,MHz to 260\,MHz. During data reduction, this allowed sideband deconvolution and the identification of any ``false'' signal due to a strong feature in the other sideband, which might otherwise be folded over onto the weak \molo\ feature. 

Both the Wide Band Spectrometer (WBS) and the High Resolution Spectrometer (HRS) with respective resolutions 1.10\,MHz and 250\,kHz (0.68 and 0.15\,\kms) were used. The positions of spikes and spectrometer artifacts are known and flagged and none of these interfered with the \molo\ observations. Initial data reduction and calibrations were performed with the standard {\it Herschel} pipeline HIPE version 7.1 and the subsequent data analysis was made with a variety of other software packages. The quality of the HIFI data is excellent and it was sufficient to subtract a low order baseline for each LO setting prior to averaging. The intensity calibration accuracy is estimated to be 10\%.

The frequency dependent beam widths (HPBW) and main beam efficiencies ($\eta_{\rm mb}$) were adopted from \citet{roelfsema2012}.  Before averaging the data for a given position, the H- and V-polarization spectra were all inspected individually in order to avoid the suppression of weak spectral lines or the generation of artificial features.

\subsection{774\,GHz observations}

The 774\,GHz observations with HIFI Band\,2 were made on OD\,583 (2011 September 15). A $6\times 6$ map, aligned with the equatorial celestial coordinates (Fig.\,\ref{774_obs}), was obtained with 10\asec\ regular spacing, oversampling the 28\asec\ beam by nearly a factor of three and  allowing for potentially increased spatial information of the emission regions. 

Both the WBS (0.43\,\kms) and the HRS (0.097\,\kms) were used. The 774\,GHz data were also obtained in DBS mode, with the offset positions displaced by 3\amin\ and in nearly the same chopping direction as before (along $pa = 98$\adeg). With 500\,s on-source integrations at each position, the total observing time was 12\,hr.

\section{Results}

In the figures showing HIFI data, brightness is given as antenna temperture, \ta. When more than one telescope had been  used, the data are presented on the main beam brightness temperature scale, \tmb, where the main beam efficiency is $\eta_{\rm mb}=0.75$ for both HIFI transitions \citep{roelfsema2012} and 0.9 for the Odin transition \citep{frisk2003,hjalmarson2003} and where we use the relation \ta/$\eta_{\rm mb}=$\,\tmb. For completeness, we also provide in Table\,\ref{results}, together with the HIFI data, the results of the Odin observations. 

\subsection{487\,GHz}

The reduced WBS spectra for the three positions O1, O3 and O4 are shown for the frequency range 484.65\,GHz to 489.85\,GHz in Fig.\,\ref{obs}a. The Upper Side Band (USB) is completely line-free, i.e. all line detections refer to the LSB only. The presented data are DSB in order to suppress the noise. Centered on the \molo\ line, radial velocities span --1600 to +1600\,\kms. In addition to the detected \molo\ 487.25\,GHz lines near an LSR velocity of 3\,\kms\ in all three positions, the spectra show a number of other emission features which correspond to identified molecular lines, due to, e.g., CS, CH$_3$OH, NH$_2$D and SO, to mention only the strongest ones. Also these spectra reveal variations in intensity and line width on angular scales of only some tens of arcsec. This remarkable behaviour of  different molecular species toward the \roac\ has been known for some time, and has recently been documented by \citet{bergman2011a}. Our multi-line HIFI data will not be discussed further here but be presented elsewhere (Larsson et al., Pagani et al., in preparation).

Figure\,\ref{obs}b shows the WBS spectra zoomed in on the \molo\,487\,GHz transition. This line too displays varying intensity levels in the three positions. In particular, the line toward the O1 position appears, by comparison, rather strong and broader than those toward O3 and O4. The HRS data shown in Fig.\,\ref{obs}c suggest the line at O1 to be a composite of two different, narrow components, blending into a single broad feature at the lower resolution of the WBS. Henceforth we will refer to this spectral component at \vlsr\,$\sim +3.6$\,\kms\ as O1$\_$2, whereas the one  closer to 2.8\,\kms\ is referred to as O1$\_$1.

We summarise the spectral characteristics of the \molo\ 487\,GHz line in Table\,\ref{results}. Within the errors, the WBS line centroids and widths are in agreement with the Gaussian fittings of the HRS data. Also, the average intensities for positions O1$\_$2, O3 and O4 with the WBS ($8.3\pm 1.8$\,mK \kms) and the HRS ($9.1\pm 1.9$\,mK \kms), respectively, are consistent with each other.

In addition, the \molo\ line centroid and width also conform with values of other optically thin species, e.g. $^{13}{\rm C}^{18}{\rm O}\,(3-2)$ \citep{liseau2010}, N$_2$H$^+$\,(1--0) \citep{diFrancesco2004}, deuterated molecules \citep{bergman2011a} and the recently discovered \water$_2$ \citep[HOOH,][]{bergman2011b}. Regarding the 2.8\,\kms\ feature, we can at present not entirely exclude the possibility that it is due to an unidentified species. If so, the line transition frequency should be $\nu_0=(487250.548\pm 0.098)$\,MHz ($1\sigma$).

\begin{figure}[t]
  \resizebox{\hsize}{!}{
  \rotatebox{00}{\includegraphics{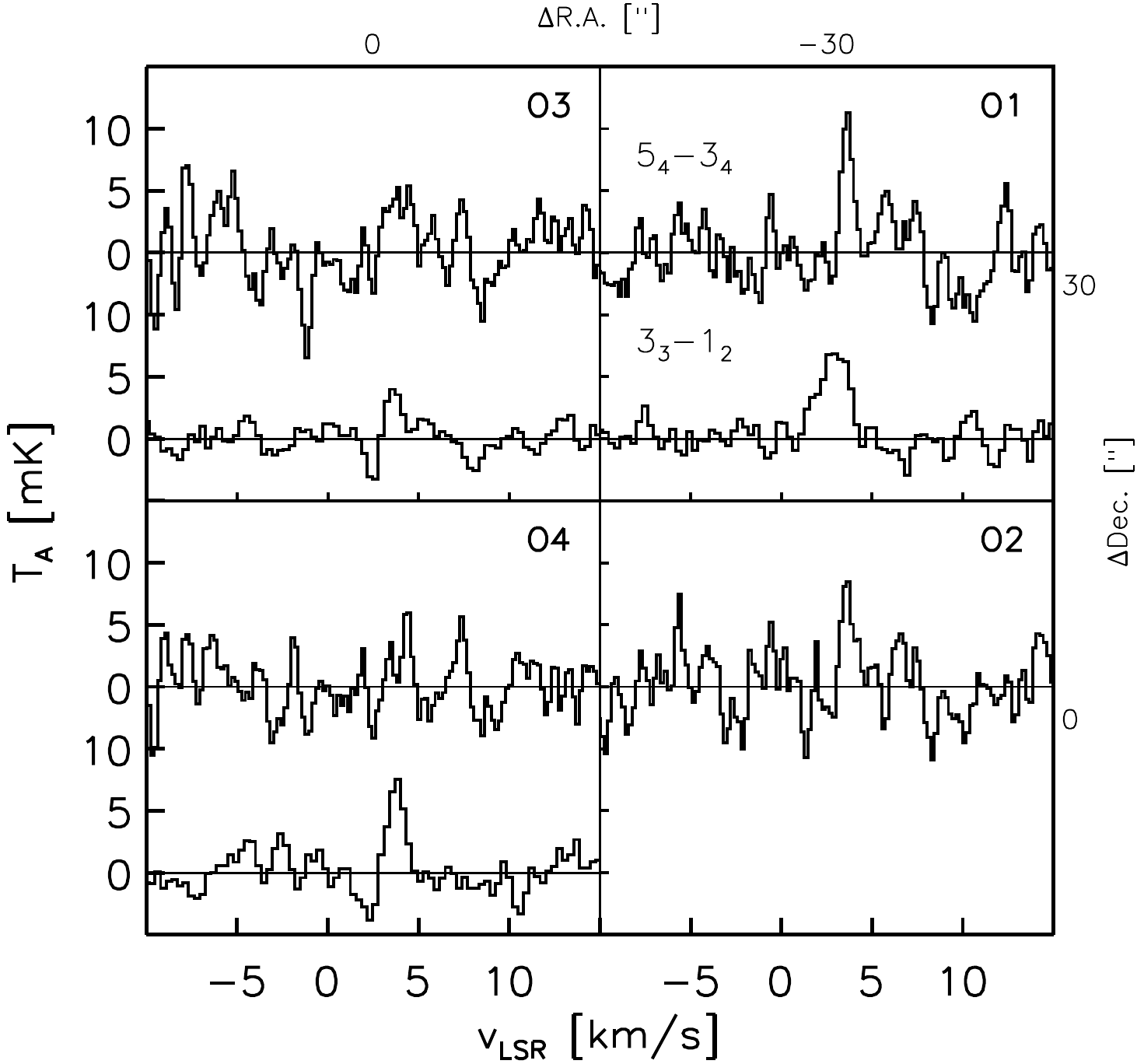}}
                        }
  \caption{The WBS spectra toward the positions O1 to O4 are shown for both the 774\,GHz (upper) and 487\,GHz (lower) lines in each frame. The averaged 774\,GHz data are convolved to the 487\,GHz beam of 44\asec. Offsets relative to the origin O4 are shown along the upper and right-hand scales.}
  \label{487_774}
\end{figure}

\subsection{774\,GHz}

The convolved and averaged 774\,GHz spectra are compared to 487\,GHz spectra in Fig.\,\ref{487_774} and the $6 \times 6$ map of the 774\,GHz observations, convolved to the beam at 487\,GHz, is displayed in Fig.\,\ref{774_map}. The line was not detected toward the east side of the cloud, i.e. toward  O3 and O4, with $1 \sigma$-upper limits of 3\,mK\,\kms, but clearly detected toward the west side of the cloud, i.e. toward O1 and O2, at $S/N=  4 - 6$. The derived line parameters for the 774\,GHz spectra, convolved to the 487\,GHz beam, are reported in Table\,\ref{results}.  We note that only the 3.6\,\kms\ component is seen at the O1 position. 

The \molo\,774\,GHz line falls in the lower side band (LSB: 770.75--775.65\,GHz). The entire observed spectral region is however totally dominated by the $^{13}$CO\,(7--6) transition at 771.2\,GHz. This line exhibits significant variations on small angular scales which is clearly evident in individual pointings of the 774 GHz beam of 28\asec\ (FWHM). In the spectrum farthest to the northeast,  the line reaches a peak \tmb\ of 25\,K, whereas in the opposite corner in the southwest the intensity has decreased to less than 5\,K.

In addition, in the upper side band (USB: 783.15--786.815\,GHz), the strongest and clearly detected line is due to the C$^{17}$O\,(7--6) transition at 786.3\,GHz (Larsson et al., in preparation). In contrast to $^{13}$CO, the line of C$^{17}$O has its maximum of almost 3\,K at the O1 position. 

\section{Discussion}

\subsection{Observed quantities: Temperature, line width, column density and size of the \molo\ emission region }

The \molo\ 487\,GHz and 774\,GHz lines are clearly detected toward multiple positions. From the observation of more than one transition, some basic physical parameters of the source can be derived, where we assume that all transitions trace the same gas along the line of sight. In the case of the observed \molo\ emission, this should be straightforward, as detailed computations of statistical equilibrium and radiative transfer for a multi-level system demonstrate this emission to be optically thin and to originate in conditions of local thermodynamic equilibrium (LTE).

\subsubsection{Temperature}

\begin{figure}[ht]
  \resizebox{\hsize}{!}{
  \rotatebox{00}{\includegraphics{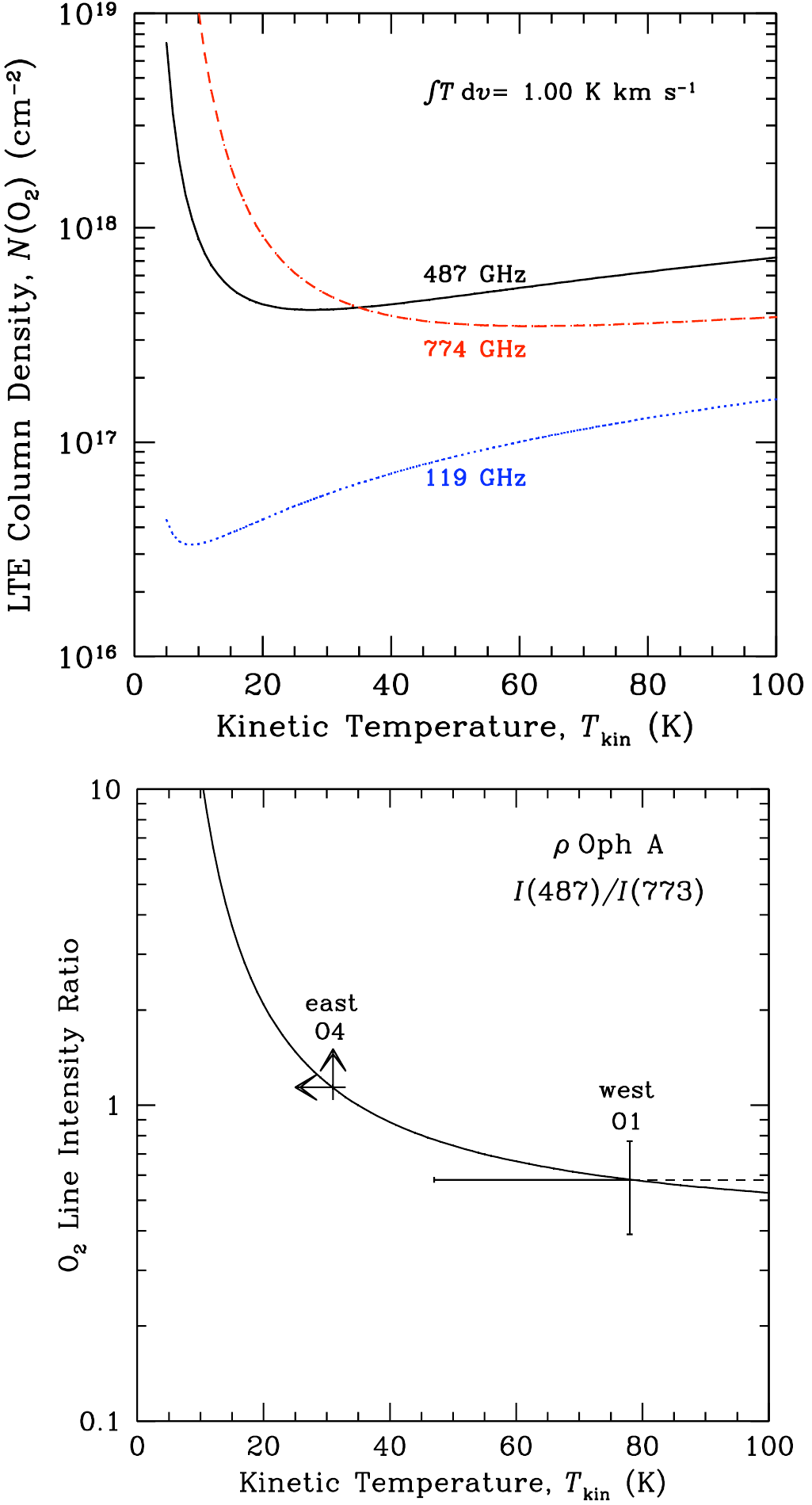}}
                        }
  \caption{{\bf Upper:} LTE computations of the \molo\ column density $N$(\molo) as function of the temperature $T$ for an extended homogeneous source, to yield a line intensity of 1.0\,K\,\kms. For optically thin emission, intensity and column density are linearly related. 119\,GHz: blue dots; 487\,Ghz: black solid line; 774\,GHz: red short dashes. {\bf Lower:} The ratio of the 487\,GHz integrated line intensity to that at 774\,GHz is shown  by the curve for a range of temperatures. For O1 (= O1$\_$2), the error shown by the dashed line formally extends to the unplausible result of nearly \powten{3}\,K. For O4, the limit symbols are for the lower limit on the intensity ratio and the upper limit on the temperature, respectively.}
  \label{o2_composite}
\end{figure}

At O1, the observed source extent of the 487\,GHz and 774\,GHz lines seems similar so that the ratio of the beam fillings $f_{487}/f_{774}$ is probably not far from unity. In addition, the main beam efficiencies at these  frequencies are also about equal  (Table\,\ref{lines}). Therefore, the observed ratio of integrated intensities  $I(487)/I(774)_{\rm obs}$ is directly comparable to the theoretical one,  $I(487)/I(774)_{\rm theo}$, obtained from LTE calculations for optically thin \molo\ radiation and which as such depends only on the gas temperature (Fig.\,\ref{o2_composite}). 

From Table\,\ref{ratios}, we see that the measured intensity ratio for the 487\,GHz and 774\,GHz lines toward O1 implies a temperature of nearly 80\,K, with the large error allowing a large range, from about 50\,K to very much higher than 100\,K (see also Fig.\,\ref{o2_composite}). If the beam filling at 774\,GHz were smaller than that at 487\,GHz, the actual temperatures would be even higher. Our value can be compared to earlier estimates, obtained at comparable spatial resolution, of both the dust and the gas temperature toward O1. From FIR continuum measurements with a 40\asec\ beam, \citet{harvey1979} obtained the dust temperature $T_{\rm d}=41$\,K. Also with a 40\asec\ beam, \citet{zeng1984} derived from NH$_3$ observations the temperature of the gas, $T_{\rm g}\sim 45$\,K.

For the dense spot O4 (P\,3, SM\,1), the $3\sigma$ upper limit on the 774\,GHz line yields a temperature strictly lower  than $31$\,K,which would be in agreement with $T_{\rm g}=22 \pm 3$\,K obtained by \citet{bergman2011b}. For the dense core \roa, the average for $<$O1, O3, O4$>$ results in a value of about 20 to 25\,K, which is entirely consistent with earlier determinations, based on a variety of techniques. Temperatures between 9\,K and 49\,K have been derived, but with most values clustering around 20--30\,K \citep[e.g.,][]{harvey1979,loren1980,ward-thompson1989,motte1998,stamatellos2007,bergman2011a,ade2011}.  

\begin{figure*}[ht]
  \resizebox{\hsize}{!}{
  \rotatebox{0}{\includegraphics{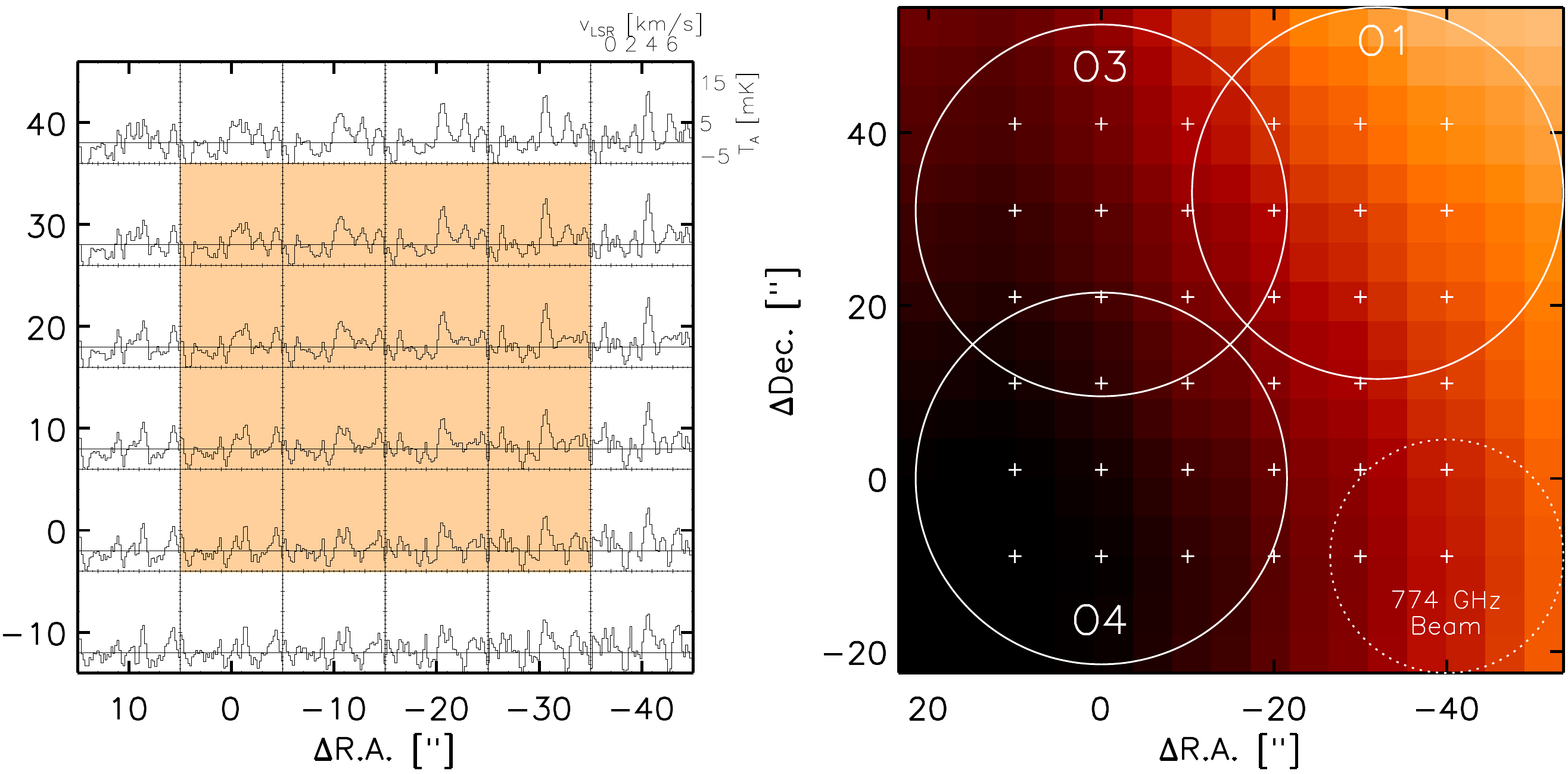}}
                        }
  \caption{{\bf Left:}  The regridded $6 \times 6$ map of the WBS 774\,GHz spectra,  weighted by the Gaussian profile of the 487\,GHz beam of 44\asec. The darker area shows roughly the region inside that Half Power Beam Width (HPBW). The \vlsr\ and \ta\ scales are indicated in the upper right corner. 
  {\bf Right:} Same as to the left but shown as an image of the spatial distribution of the integrated intensity, \tadv, with observed positions shown as plusses. The circles show the 487\,GHz and 774\,GHz beams, respectively.
  }
  \label{774_map}
\end{figure*}

As is also evident from Fig.\,\ref{774_map}, a temperature gradient or two distinctly different temperature regimes, i.e. with larger than about $50$\,K and strictly lower than $30$\,K, exist within the boundaries of our limited map of roughly one square arcminute (see also Figs.\,\ref{dbs} and \ref{774_obs}). 

\begin{table}
\caption{Observed \molo\, line intensity ratios. }        
\label{ratios}      
\begin{tabular}{ccrc}      
\hline\hline    
       \noalign{\smallskip}   
Spectrum/ 	& \underline{$I(487)^{a,\,b}$} &	$T_{\rm kin}$ 		&	$N$\,(\molo)	 	 \\
Position 		& $I(774)$		    	&	(K)				&	(\cmtwo)			\\	
\hline	     
\noalign{\smallskip}   	
O1$\_1$ (2.8\,\kms)	& $>0.9\,(3 \sigma)$	&  $< 40$				& $> 2 \times 10^{15}$      \\
\noalign{\smallskip}    
O1$\_2$ (3.6\,\kms)	& $0.58(\pm 0.19)$	&  $78$				& $5.4\times 10^{15}$ 	  \\
\noalign{\smallskip}    
			&($+ 1 \sigma$) 0.77	&  $47$				& $5.0\times 10^{15}$	 \\
\noalign{\smallskip}    
			& ($-1 \sigma $) 0.39	&  $\gg100$			& $\dots$	  			\\		
\noalign{\smallskip}   
O3			& $>0.61\,(3\sigma)$	& $< 70$				& $>3\,\times 10^{15}$	\\
\noalign{\smallskip}  
O4			& $>1.14\,(3\sigma)$	& $< 31$				& $> 6 \times 10^{15}$	\\
\noalign{\smallskip}   
\noalign{\smallskip}
$<$O1, O3, O4$>$& $1.43(\pm 0.94)^c$	&  $25$				& $3.4\times 10^{15}$	  \\
\noalign{\smallskip}   
			&($+ 1 \sigma$) 2.37	&  $18$				& $4.6\times 10^{15}$	 \\
\noalign{\smallskip}    
			& ($-1 \sigma $) 0.49	&  $>100$				& $\dots$	 			 \\		

\hline	     
  \noalign{\smallskip}
  \hline
   \noalign{\smallskip}  
\end{tabular}
\begin{list}{}{}
\item[$^{a}$] Errors are $\pm (I_1/I_2) \sqrt{(\delta I_1/I_1)^2 + (\delta I_2/I_2)^2}$.
\item[$^{b}$] For HIFI, $\eta_{\rm mb}(487)=\eta_{\rm mb}(737)=0.75$.
\item[$^{c}$] Similar to O3 for $1\sigma$ limit on 774\,GHz intensity ($1.47\pm 1.26$).
\end{list}
\end{table}

\subsubsection{Line width}

Comparing the observed line widths, FWHM, of Table\,\ref{results} with the resolutions offered by the HRS and WBS, we see that the HRS data for the 487\,GHz line are all well resolved ($\delta \upsilon = 0.15$\,\kms), whereas the 774\,GHz line toward O1 and O2 is just barely resolved with the WBS  ($\delta \upsilon = 0.43$\,\kms). However, the average 774\,GHz spectrum $<$O1, O2$>$ is well resolved with the HRS and the line has a comparable width to what is observed with the WBS. What is remarkable is that the lines in the cold regions O3 and O4 are observed to be broader than those in the warmer O1 and  O2. 

The non-thermal contribution to the observed width of an unblended line is  $\upsilon_{\rm non-th} = ({\small {\rm FWHM}}^2 - \upsilon_{\rm th}^2)^{1/2}$, where all radial velocities are expressed in terms of their full width half maximum value. For $T$ in the range 18\,K to 30\,K, $\upsilon_{\rm th} = 0.27$ to 0.35\,\kms, yielding  $\upsilon_{\rm non-th} =0.72$ to 0.80\,\kms\ for the cold region including O3 and O4  (Table\,\ref{ratios}). Hence, we find that there the line broadening is totally dominated by the non-thermal random motions, being more than twice as high as the thermal ones. On the other hand, in the warm region O1 and O2, the thermal motions dominate over the non-thermal ones (0.57\,\kms\ versus $\le 0.44$\,\kms\ for $T=78$\,K) or being comparable at the lower limit temperature of  47\,K.  This could suggest that, if the non-thermal motions are identified with turbulence at O1 and O2, much of this turbulence has been dissipated and converted into heat. 
 
The O2 source is seen in projection close to the  outflow from VLA\,1623 and judging from its position (Figs.\,\ref{dbs} and \ref{774_obs}), it is possible that the gas at O2 is enriched by shock processed material.  At the border of the outflow, a broader line might be expected due to enhanced turbulent motions. In fact, the 774\,GHz line width appears (marginally) broader than that toward O1 (Table\,\ref{results}), but the uncertainties are large and the significance is low.
 
\subsubsection{Source size}

\begin{figure}[t]
  \resizebox{\hsize}{!}{
  \rotatebox{00}{\includegraphics{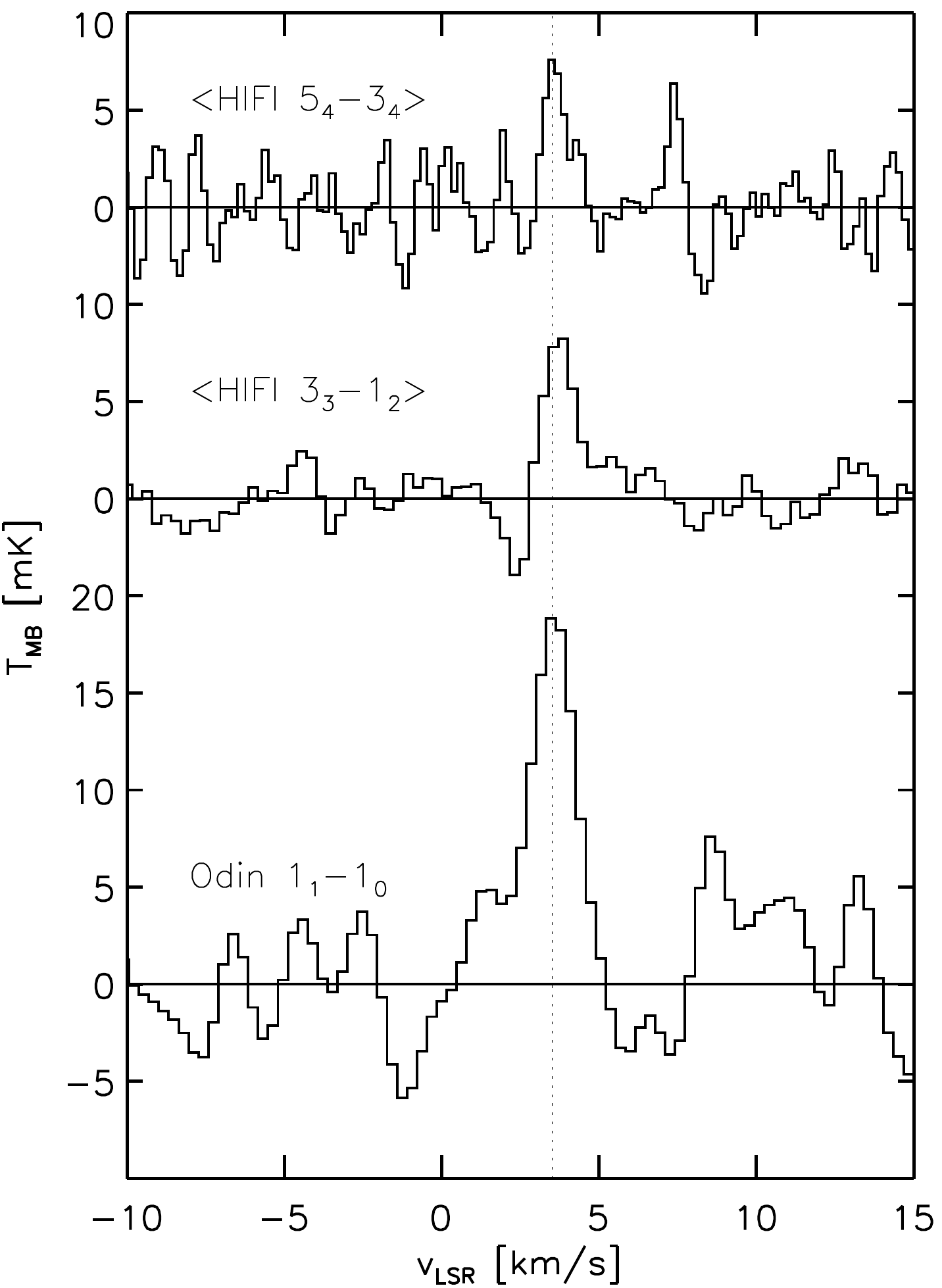}}
                        }
  \caption{Averaged data for the HIFI \molo\ observations (774 and 487\,GHz) and the Odin 119\,GHz spectrum toward \roa. For proper comparison, these spectra are given in the \tmb\ scale and for reference, a vertical dotted line is shown at \vlsr\,=\,+3.5\,\kms.}
  \label{all_spectra}
\end{figure}

Having determined the temperature, we can use the theoretical line ratio  for another transition and equate it to the ratio of the respective beam fillings, e.g.  $R \equiv (I_{119}/I_{487})_{\rm theo}(T)/(I_{119}/I_{487})_{\rm obs}=f_{487}/f_{119}$. The relevant 119\,GHz data have been adopted from \citet{larsson2007} and are compiled in Table\,\ref{results} and all lines are displayed on the \tmb\ scale in Fig.\,\ref{all_spectra}. For a Gaussian beam at 119\,GHz, the beam filling factor is given by  $f_{119} \approx \theta_{119}^2/(\theta_{119}^2 + \theta_{\rm Odin}^2)$, where $\theta_{119}$ is the size of the source and $\theta_{\rm Odin}$ that of the beam at 119\,GHz. Re-arranging, this reads $\theta_{119} \approx  \theta_{\rm Odin}/\sqrt{1/f_{119} - 1} = \theta_{\rm Odin}/\sqrt{R/f_{487} - 1}$, where $\theta_{\rm Odin}=10$\amin. 

The 487\,GHz source seems to be extended on the scale of at least the beam width, i.e. $f_{487}\ge 0.5$, and for the average values of $<$O1, O3, O4 $>$ in Table\,\ref{ratios}, we find a size of the Odin source of \gapprox\,4 to 5\,arcmin, for gas at 18 to 25\,K, i.e. for $(I_{119}/I_{487})_{\rm theo}(T) = 11$ to 8, respectively. The solution for a much warmer gas (127\,K for ``$-1 \sigma$" in Table\,\ref{ratios}) ought to be excluded, as the calculated source size would exceed 9\amin, whereas  the high temperature tracing 774\,GHz line has been detected only over a limited region (\lapprox\,1\amin). On the basis of CS\,(2-1) mapping observations, \citet{liseau1995} determined  the deconvolved angular size of the densest regions of the \roac\ as FWHM\,\about\,4\amin\ ($n_{\rm H}$\,\gapprox\,$10^{5}$\,\cmthree), a \roa\ size consistent with the HIFI and Odin observations of \molo. It appears that the entire \roac\ is faintly glowing in the 119\,GHz line.

\subsubsection{Column density}

For the warm region O1, an \molo\ column density of $5 \times 10^{15}$\,\cmtwo\ is derived, whereas for the cold O4, $N$(\molo)\,$> 6 \times 10^{15}$\,\cmtwo\ (Table\,\ref{ratios}). These estimations are based on the assumption of optically thin emission in LTE at the kinetic gas temperature $T_{\rm kin}$, i.e. for unit beam filling the \molo\ column is given by
\\

$
N({\rm O}_2) =   \int\!T_{\rm mb}\,{\rm d}\upsilon \times \,\Phi(T_{\rm kin}) \,\,\,{\rm cm}^{-2}
$
\\

\noindent
when the integral is expressed in K\,\cms\ and where
\\

$
\Phi(T_{\rm kin}) =  \left ( \frac {2 \pi^{1/3} k}{h c} \right )^3 \frac {T^2_{\rm tr}} {g_{\rm ul} A_{\rm ul}} \,\, \frac {F_{\nu}(T_{\rm kin})}{F_{\nu}(T_{\rm kin}) - F_{\nu}(T_{\rm bg})} \,\, Q(T_{\rm kin}) \,\exp({{T_{\rm up}}/T_{\rm kin}}),
$
\\

\noindent
in K$^{-1}$\,\cmthree\,s (cgs units throughout). Here, the transition temperature is $T_{\rm tr} \equiv h \nu/k$, the quasi-Planck function is $F_{\nu}(T) \equiv T_{\rm tr}/(e^{{T_{\rm tr}/T}}-1)$, the partition function is $Q(T)$, the upper level energy is $T_{\rm up}=E_{\rm up}/k$ and the other symbols have their usual meaning. The background temperature is that of the cosmic microwave background, i.e. $T_{\rm bg}=T_{\rm CMB}$.\footnote{In our numerical radiative transfer calculations (Sect.\,4.1), also the dust continuum is included. These calculations make use of the collision data of \citet{lique2010}.} The frequencies were adopted from \citet{drouin2010} and the Einstein $A$-values from \citet{marechal1997} \citep[see also][]{blacksmith1984}.

If on the other hand $<$O1, O3, O4$>$ can be taken as a representative for \roa\ as  a whole, column densities lower than $5\times 10^{15}$\,\cmtwo\ would be indicated.  Although there is a considerable range in gas kinetic temperature, the \molo\ column densities seem not to vary much. However, with the observed inhomogeneity of the emission in  mind, the sigificance of such an average could be questioned. 

\subsection{\molo\ abundance}

Once the \molo\ column density is known, the line of sight average \molo\ fractional abundance, with respect to \molh\ or H-nuclei, can be measured if the column density of \molh\ molecules is available. The hydrogen column density along the  appropriate lateral and depth scales is not well known from observation.  However, we can get an approximate idea from the C$^{18}$O observations by \cite{liseau2010}, where it was suggested on the basis of the line profiles that the $J$=3--2 line was a reasonably good proxy for the \molo\,119\,GHz line observed with Odin \citep{larsson2007}. If the \molo\ molecules are indeed cospatial with those of CO and its isotopic variants, their optically thin lines could be used to infer the \molh\ column densities relevant also for \molo. We derive ``local" C$^{18}$O column densities taking C$^{18}$O intensity and \molo\ temperature variations into account. The gas and dust peaks coincide spatially (see also Table\,\ref{sources}), i.e. there is no indication of significant CO freeze 
out \citep[at levels higher than a factor of 2 to 3, see Fig.\,10 of][]{bergman2011a}.  The C$^{18}$O column density varies by less than 20\%  for $20\,{\rm K} \le T_{\rm kin}\le 50$\,K. Assuming that C$^{18}$O/\molh\,$= 1.5 \times 10^{-7}$ \citep[likely within about 30\%, e.g.][]{wannier1976,goldsmith2000}, \molh\ columns are found to be $1 \times 10^{23}$\,\cmtwo. 

Hence, for the \molo\ column densities in Table\,\ref{ratios}, we find that $X$(\molo)\,$ = 5 \times 10^{-8}$ relative to \molh\  in the warm gas toward the west, at O1 (and likely also toward O2), and higher than that toward the east, at the colder O4. The C$^{18}$O lines are slightly optically thick, implying that these \molh\ column densities are lower limits and need to be adjusted upwards by a factor of the order of  $\tau/(1-e^{-\tau})$\,\gapprox\,2, for $\tau \sim 2$. Hence, these \molo\ abundances are upper limits, i.e. the fractional abundance of \molo\  in the gas phase is less than indicated above. 

A way to obtain a more detailed picture of the source is to use detailed theoretical models to predict the \molo\ abundance or column and to compare these models with the observations. This will be the topic of the next section.

\subsection{Chemical models}

Specifically for  \molo\ (and \water), \citet{hollenbach2009}  constructed theoretical models of Photon-Dominated Regions \citep[PDRs,][]{hollentiel1997}, invoking also grain surface chemistry, including freeze-out and desorption of the species. These new models are models of the whole cloud, to arbitrary values of the extinction, and showing that most of the gas phase \molo, OH and \water\ is found at values of \av\,\about\,3--10\,mag, where the details depend, as in a conventional PDR, on the ratio $G_0/n_{\rm H}$. In many aspects, these models are similar to the traditional dense PDR models including chemistry and gas heating which have been developed to explain the abundances and column densities of other molecules (including radicals and ions) at intermediate depths into the clouds \citep[e.g.,][]{fuente1993,sternberg1995,jansen1995,simon1997,pety2005}. The current models are extended even deeper into the cloud where external UV photons play no role anymore in the chemistry, only cosmic-ray induced photons.

In the direction of \roa, a number of spectroscopic indicators suggest the presence of a PDR\footnote{Atomic recombination lines of carbon, $^{12}$C, and sulfur, $^{32}$S, \citep{cesarsky1976,pankonin1978}, fine structure lines:  [C\,I]\,492.2\,GHz, \citep{kamegai2003}, [C\,I]\,809.3\,GHz \citep{kamegai2003,kulesa2005},  [C\,II]\,157\,\um\ \citep{yui1993,liseau1999},  [O\,I]\,63\,\um\ \citep{ceccarelli1997,liseau1999}, [O\,I]\,145\,\um\ \citep{liseau1999,liseaujusttanont2009} and mid-infrared PAH emission: \citep{liseaujusttanont2009}. }. \citet{liseau1999} inferred that the PDR is situated on the rear side of \roa, as seen from the Sun, and excited by the two stellar sources HD\,147889 and S\,1 (see Fig.\,\ref{dbs} and Table\,\ref{sources}) by their combined UV fields, corresponding to $G_0 \sim 10^2$.  For the densest parts of the [C\,II] emitting \roa-PDR these authors derived empirically $n_{\rm H}=5 \times 10^4$\,\cmthree\ and over more extended regions,  1 to 3\,$\times 10^4$\,\cmthree.

The model of \citet{hollenbach2009} predicts columns of \molo\ between about $10^{15}$ and a few times $10^{16}$\,\cmtwo\ depending on the grain temperature in the \molo-emitting region, with higher columns in regions with warmer dust.   Warmer ($T_{\rm d} > 20$\,K) dust produces more \molo\ column because O atoms evaporate from grains before they can react with atomic hydrogen sticking to the grains. This increases the elemental gas phase O available to make \molo, whose abundance in the \ro\ model can reach locally above \powten{-5} for $T_{\rm d}=25$\,K. The results are quite insensitive to the gas temperature in the \molo-emitting region for the range 10\,K$ <T_{\rm g} <100$\,K.  

However, the \citet{hollenbach2009} models predict  gas temperatures of only $\sim 20$\,K, as would all thermochemical models with only cosmic ray heating and heating by the penetrating FUV.  Gas temperatures as high as 80\,K, as derived here,  suggest other heating mechanisms, such as slow shocks or turbulent dissipation \citep[e.g.,][]{goldsmith2010}. Our observations show a slight increase in \molo\ column in {\it cooler} gas. If the dust is also cooler in the cooler gas region, as would be expected for the derived  densities above $10^4$ and, centrally, higher than $10^7$\,\cmthree, this contradicts the predictions of this particular model.  

On the other hand, geometry also plays a role in determining the column of \molo\ observed along the line of sight.  The \citet{hollenbach2009}  models assume an illuminated slab face-on.  However, clumpiness and more edge-on geometry could increase the observed column in some beams, and could explain the slightly increased column of \molo\ in the region with somewhat cooler gas.   We conclude that these models are modestly consistent with the observations, but likely require some additional gas heating in the \molo-emitting region.  We defer detailed thermochemical modeling of these \ro\ observations to a future paper.

\subsection{Source(s) of \molo\  or the dearth of \molo}

The discussed models could be expected to apply quite generally to all externally illuminated molecular clouds developing PDR-like parts. It is surprising therefore that  \molo\ has been detected only toward the dense cloud core \roa\ \citep[][and this work]{larsson2007} and a spot in OMC-1 \citep{goldsmith2011}, in spite of numerous searches toward other regions. For instance, surveying twenty sources, \citet{goldsmith2000} obtained limiting \molo\ abundances $N$(\molo)/$N$(\molh) of a few times \powten{-7}, whereas \citet{pagani2003} derived limiting \molo\ columns of  $10^{15}$ to a few times $10^{16}$\,\cmtwo\  for nearly a dozen objects.  These surveys included dark clouds, regions of low mass star formation in the solar neighbourhood and of high mass star formation at different galactocentric distances, i.e. these surveys sampled a wide range of cloud properties and energetics, from quiescent cold clouds to  turbulent shock heated gas due to mass infall and molecular outflows. Yet, most escaped detection, although their limiting \molo\ column densities fall within the range predicted by the theoretical models. 

One possibility could be that this apparent mismatch is due to observational bias, such as source distance and beam dilution. For instance, the limit on the \molo\ column in Ori\,A determined by \citet{pagani2003} is entirely consistent with the HIFI beam-averaged detection obtained by \citet{goldsmith2011}, if proper correction for  dilution in the Odin-beam is applied. We recall that the detected spot is not toward the Orion-PDR \citep{melnick2012}, as one might have expected, but close to the outflow in the vicinity of IRc\,2.

However, the difference in distance aside, with differing $G_0/n_{\rm H}$, \roa\ and Ori\,A represent very different environments, viz. cold core deuteration versus hot core grain chemistry
\citep{bergman2011a,bisschop2007,ceccarelli2007,ratajczak2011}. Star forming regions like IRAS\,16293 in Ophiuchus, NGC\,1333 in Perseus or the Serpens core near S\,68 are more akin to the conditions in \roa, yet no \molo\ detections have been reported. These complexes are at the same or within at most twice the distance to \roa, so that beam dilution issues should be much reduced. It is conceivable that the viewing geometry of \roa\ could be particularly favourable for the observability of \molo\ and that this aspect is not shared by other regions.

However, a clear difference is suggested by the fact that hydrogen peroxide (H$_2$O$_2$) has so far only been detected in \roa\ \citep{bergman2011b}. H$_2$O$_2$ is thought to be a clear indicator of grain surface chemistry and relatively abundant at about 30\,K \citep[e.g.,][]{cazaux2010}. The simultaneously detectable abundance of \molo\ and H$_2$O$_2$ may perhaps be used to infer the evolutionary chemical state of the cloud, hence its ``age", if interstellar gas phase \molo\ is relatively transient \citep{du2012}. In that case, it could explain the paucity of known \molo\ line emitters.

\section{Conclusions}

   \begin{itemize}
      \item[$\bullet$] We have successfully detected the $3_3-1_2$  transition of \molo\  at 487\,GHz with HIFI onboard {\it Herschel} toward three positions in the \roac.  In addition, we have also obtained a $6 \times 6$ oversampled map with 10\asec\ spacings in the $5_4-3_4$ transition at 774\,GHz. 
      \item[$\bullet$] The telescope pointings were toward two cold, high density cores, i.e. O4 [C$^{18}$O\,P3 = SM\,1] and O3 [C$^{18}$O\,P2, near SM\,1N], and two positions 30\asec\ west of these, labelled O2 and O1, respectively. All three observed positions (O1, O3 and O4) were detected in the 487\,GHz line, with significant variations of the line intensity on scales as small as 30\asec. At 774\,GHz, emission was detected from only the western part of the map, including O1 and O2. 
      \item[$\bullet$]  Combining the 487\,GHz and 774\,GHz data leads to the column density of $N$(\molo)\,$>6 \times 10^{15}$\,\cmtwo, at the $3 \sigma$ level and at $T < 30$\,K, in the high density region O3 and O4, whereas in the warmer region of O1 and O2, $N$(\molo)\,$=5.5 \times 10^{15}$\,\cmtwo\ ($T > 50$\,K). There, standard analysis yields an abundance of $N$(\molo)/$N$(\molh)\,$\sim 5 \times 10^{-8}$ in the warm gas and somewhat higher in the cold region.
      \item[$\bullet$] This result is in agreement with $X$(\molo) based on the observation of the 119\,GHz line with Odin \citep{larsson2007}, assuming that the \molo\ source size has an angular extent of about 5\amin.
      \item[$\bullet$] Comparing our results with the predictions from theoretical chemical models, we find good agreement between observed and predicted \molo\ column densities. Less convincing agreement regards theoretical and observed temperatures, the reasons for which are not clear.
      \item[$\bullet$] The question why \molo\ is such an elusive molecule in the ISM is still unanswered. In the special case of \roa, there is some evidence suggesting that detectable amounts of gas phase \molo\ might be a relatively transient phenomenon, which could explain why interstellar \molo\ is generally not detected.
     \end{itemize}

\begin{acknowledgements}
   We appreciate the thoughtful comments by the referee, Prof. Karl Menten, which have lead to an improvement of the manuscript. The Swedish  authors are indebted to the Swedish National Space Board (SNSB) for its continued support. This work was also carried out in part at the Jet Propulsion Laboratory, California Institute of Technology, under contract with the National Aeronautics and Space Administration (NASA).  We also wish to thank the HIFI-ICC for its excellent support. HIFI has been designed and built by a consortium of institutes and university departments from across Europe, Canada and the US under the leadership of SRON Netherlands Institute for Space Research, Groningen, The Netherlands with major contributions from Germany, France and the US. Consortium members are: Canada: CSA, U. Waterloo; France: CESR, LAB, LERMA, IRAM; Germany: KOSMA, MPIfR, MPS; Ireland, NUI Maynooth; Italy: ASI, IFSI-INAF, Arcetri-INAF; Netherlands: SRON, TUD; Poland: CAMK, CBK; Spain: Observatorio Astron—mico Nacional (IGN), Centro de Astrobiolog'a (CSIC-INTA); Sweden: Chalmers University of Technology Ð MC2, RSS \& GARD, Onsala Space Observatory, Swedish National Space Board, Stockholm University Ð Stockholm Observatory; Switzerland: ETH Z\"urich, FHNW; USA: Caltech, JPL, NHSC.
\end{acknowledgements}


\begin{thebibliography}{}
\bibitem[Ade et al.(2011)]{ade2011} Ade P.A.R., Aghanim N., Arnaud M., et al., 2011, A\&A, 536, A20
\bibitem[Andr\'e et al.(1990)]{andre1990} Andr\'e P., Martin-Pintado J., Despois D. \& Montmerle T., 1990, A\&A, 236, 180
\bibitem[Bergman et al.(2011a)] {bergman2011a} Bergman P., Parise, B., Liseau R. \& Larsson B., 2011\,a, A\&A, 527, A39
\bibitem[Bergman et al.(2011b)] {bergman2011b} Bergman P., Parise B., Liseau, R., et al., 2011\,b, A\&A, 531, L\,8
\bibitem[Bisschop et al.(2007)]{bisschop2007} Bisschop S.E., J\o rgensen J.K., van Dishoeck E.F. \& de Wachter E.B.M., 2007, A\&A, 465, 913
\bibitem[Black \& Smith(1984)]{blacksmith1984} Black J.H. \& Smith P.L., 1984, ApJ, 277, 562
\bibitem[Caratti o Garatti et al.(2006)]{caratti2006} Caratti o Garatti A., Giannini T., Nisini B. \& Lorenzetti D., 2006, A\&A, 449, 1077
\bibitem[Cazaux et al.(2010)]{cazaux2010} Cazaux S., Cobut V., Marseille M., Spaans M., Caselli P., 2010, A\&A, 522, 74
\bibitem[Ceccarelli et al.(1997)]{ceccarelli1997} Ceccarelli C., Haas M.R., Hollenbach D.J. \& Rudolph A.L., 1997, ApJ, 476, 771
\bibitem[Ceccarelli et al.(2007)]{ceccarelli2007} Ceccarelli C., Caselli P., Herbst E., Tielens A.G.G.M. \& Caux E., 2007, 
Protostars and Planets V, B. Reipurth, D. Jewitt, and K. Keil (eds.), University of Arizona Press, Tucson, 47
\bibitem[Cesarsky et al.(1976)]{cesarsky1976} Cesarsky D.A., Encrenaz P.J., Falgarone E.G., et al., 1976, A\&A, 48, 167
\bibitem[De Graauw et al. (2010)]{degraauw2010} De Graauw Th., Helmich F.P., Phillips T.G., et al., 2010, A\&A, 518, L\,6
\bibitem[Dent et al.(1995)]{dent1995} Dent W.R.F., Matthews H.E. \& Walther D.M., 1995, MNRAS, 277, 193
\bibitem[di Francesco et al.(2004)]{diFrancesco2004} Di Francesco J., Andr\'e P. \& Myers P. C., 2004, ApJ, 617, 425
\bibitem[Du et al.(2012)] {du2012} Du F., Parise B. \& Bergman P., 2012, A\&A, 538, A91
\bibitem[Drouin et al.(2010)]{drouin2010} Drouin, Brian J.; Yu, Shanshan; Miller, Charles E. et al., 2010,JQSRT, 111, 1167
\bibitem[Elias(1978)]{elias1978} Elias J.H., 1978, ApJ, 224, 453
\bibitem[Frisk et al.(2003)]{frisk2003} Frisk U., Hagstr\"om M., Ala-Laurinaho J., et al., 2003, A\&A, 402, L\,27
\bibitem[Fuente et al.(1993)]{fuente1993} Fuente A., Martin-Pintado J., Cernicharo J., Bachiller R., 1993, A\&A, 276, 473
\bibitem[Goldsmith \& Langer(1978)] {goldsmithlanger1978} Goldsmith P.F. \& Langer W.D., 1978, ApJ, 222, 881
\bibitem[Goldsmith et al.(2000)]{goldsmith2000} Goldsmith P.F., Melnick G.J.,  Bergin E.A., et al., 2000, ApJ, 539, L\,123
\bibitem[Goldsmith et al.(2002)]{goldsmith2002} Goldsmith P.F., Li D., Bergin E.A., et al., 2002, ApJ, 576, 814
\bibitem[Goldsmith et al.(2010)]{goldsmith2010} Goldsmith P.F., Velusamy T.,  Li D. \& Langer W.D., 2010, ApJ, 715, 1370
\bibitem[Goldsmith et al.(2011)]{goldsmith2011} Goldsmith P.F., Liseau R., Bell T.A., et al., 2011, ApJ, 737, 96
\bibitem[Grasdalen et al.(1973)]{grasdalen1973} Grasdalen G.L., Strom K.M. \& Strom S.E., 1973, ApJ, 184, L\,53
\bibitem[Harvey et al.(1979)]{harvey1979} Harvey P.M., Campbell M.F. \& Hoffmann W.F., 1979, ApJ, 228, 445
\bibitem[Hincelin et al.(2011)] {hincelin2011} Hincelin U., Wakelam V., Hersant F., et al., 2011, A\&A, 530, A61 
\bibitem[Hjalmarson et al.(2003)]{hjalmarson2003} Hjalmarson \AA., Frisk U., Olberg M., et al., 2003, A\&A, 402, L\,39
\bibitem[Hollenbach \& Tielens(1997)]{hollentiel1997} Hollenbach D.J. \& Tielens A.G.G.M., 1997, ARAA, 35, 179
\bibitem[Hollenbach et al.(2009)]{hollenbach2009} Hollenbach D., Kaufman M.J., Bergin E.A. \& Melnick G.J., 2009, ApJ, 690, 1497
\bibitem[Jansen et al.(1995)]{jansen1995} Jansen D.J., van Dishoeck E.F., Black J.H., Spaans M., Sosin C., 1995, A\&A, 302, 223
\bibitem[Johnstone et al.(2000)]{johnstone2000} Johnstone D., Wilson C.D., Moriarty-Schieven G. et al., 2000, ApJ, 545, 327
\bibitem[Kamegai et al.(2003)]{kamegai2003} Kamegai K., Ikeda M., Maezawa H., et al., 2003, ApJ, 589, 378
\bibitem[Kulesa et al.(2005)]{kulesa2005} Kulesa C.A., Hungerford A.L., Walker C.K., et al., 2005, ApJ, 625, 194
\bibitem[Larsson et al.(2007)] {larsson2007} Larsson B., Liseau R., Pagani L. et al., 2007, A\&A, 466, 999
\bibitem[Liseau et al.(1995)]{liseau1995} Liseau R., Lorenzetti D., Molinari S., et al., 1995, A\&A, 300, 493 
\bibitem[Liseau et al.(1999)]{liseau1999} Liseau R., White G.J., Larsson B., et al., 1999, A\&A, 344, 342
\bibitem[Liseau \& Justtanont (2009)]{liseaujusttanont2009} Liseau R. \& Justtanont K., 2009, A\&A, 499, 799
\bibitem[Liseau et al.(2010)] {liseau2010} Liseau R., Larsson B., Bergman P. et al.2010, A\&A, 510, 98
\bibitem[Lique(2010)]{lique2010} Lique F., 2010, J. Chem. Phys., 132, 04431
\bibitem[Loinard et al.(2008)]{loinard2008} Loinard L., Torres R.M.,  Mioduszewski A.J. \& Rodr'guez L.F., 2008, ApJ, 675, L\,29
\bibitem[Lombardi et al.(2008)]{lombardi2008} Lombardi M., Lada C. J. \& Alves J., 2008, A\&A, 480, 785
\bibitem[Loren et al.(1980)]{loren1980} Loren R.B., Wootten A. \& Wilking B.A., 1990, ApJ, 365, 269	
\bibitem[Mamajek(2008)]{mamajek2008} Mamajek E.E., 2008, Astron.Nachr., 329, 10
\bibitem[Mar\'echal et al.(1997)]{marechal1997} Mar\'echal P., Viala Y.P. \& Benayoun J.J., 1997, A\&A, 324, 221
\bibitem[Melnick et al.(2000)]{melnick2000} Melnick G.J., Stauffer J.R., Ashby M.L.N., et al., ApJ, 539, L\,77
\bibitem[Melnick et al.(2012)]{melnick2012} Melnick G.J., et al., 2012, preprint 
\bibitem[Motte et al.(1998)]{motte1998} Motte F., Andr\'e P. \& Neri R., 1998, A\&A, 336, 150
\bibitem[Nordh et. al(2003)]{nordh2003} Nord H.L., von Sch\'eele F., Frisk U., et al., 2003, A\&A, 402, L\,21
\bibitem[Pagani et al.(2003)]{pagani2003} Pagani L., Olofsson A.O.H., Bergman P., et al. 2003, A\&A, 402, L77
\bibitem[Pankonin  \& Walmsley (1978)]{pankonin1978} Pankonin V. \& Walmsley C.M., 1978, A\&A, 64, 333
\bibitem[Pety et al.(2005)]{pety2005} Pety J., Teyssier D., Foss\'e D., et al., 2005, A\&A, 435, 885
\bibitem[Pilbratt et al.(2010)]{pilbratt2010} Pilbratt G.L., Riedinger J.R., Passvogel T.,  et al., 2010, A\&A, 518, L\,1
\bibitem[Ratajczak et al.(2011)]{ratajczak2011} Ratajczak A., Taquet V., Kahane C., et al., 2011, A\&A, 528, L\,13
\bibitem[Roelfsema et al.(2012)]{roelfsema2012} Roelfsema P.R., Helmich F.P., Teyssier D., et al., 2012, A\&A, 537, 17
\bibitem[Simon et al.(1997)]{simon1997} Simon R., Stutzki J., Sternberg A. \& Winnewisser G., 1997, A\&A, 327, L\,9
\bibitem[Snow et al.(2008)]{snow2008} Snow T.P.,  Destree J. \& Welty D.E.,  2008, ApJ, 679, 512
\bibitem[Stamatellos et al.(2007)]{stamatellos2007} Stamatellos D., Whitworth A.P. \& Ward-Thompson D., 2007, MNRAS, 379, 1390
\bibitem[Sternberg \& Dalgarno(1995)]{sternberg1995} Sternberg A. \& Dalgarno A., 1995, ApJ, 99 565
\bibitem[Wannier et al.(1976)]{wannier1976} Wannier P.G., Penzias A.A., Linke R.A. \& Wilson, R.W., 1976, ApJ, 204, 26
\bibitem[Ward-Thompson et al.(1989)]{ward-thompson1989} Ward-Thompson D., Robson E.I.,  Whittet D.C.B. et al., 1989, MNRAS, 241, 119
\bibitem[Yui et al.(1993)]{yui1993} Yui Y.Y., Nakagawa T., Doi Y., et al., 1993, ApJ, 419, L\,37
\bibitem[Zeng et al.(1984)]{zeng1984} Zeng Q., Batrla W. \& Wilson T.L., 1984, A\&A, 141, 127
\end{thebibliography}
\end{document}